\documentclass{article}
\usepackage{dcase2024,amsmath,graphicx,url,times,booktabs, tabularx}

\usepackage{latexsym}
\usepackage[T1]{fontenc}
\usepackage[utf8]{inputenc}
\usepackage{microtype}
\usepackage{inconsolata}
\usepackage{amsfonts}
\usepackage{multirow}
\usepackage{arydshln}
\usepackage{xspace}
\usepackage{tikz}
\usetikzlibrary{shapes.geometric, arrows, positioning, fit}
\newcommand{\etal}{\textit{et~al.}\xspace}

\title{EnCLAP++: Analyzing the EnCLAP Framework for Optimizing Automated Audio Captioning Performance}

\name{Jaeyeon Kim$^{1,2}$,
       Minjeong Jeon$^{2}$,
       Jaeyoon Jung$^{2,3}$, 
       Sang Hoon Woo$^{4}$,
       Jinjoo Lee$^{2}$, 
       }
 \address{$^1$ Seoul National University, Seoul, Republic of Korea, jaeyeonkim99@snu.ac.kr \\    
         $^2$ MAUM AI Inc.,  Seongnam, Republic of Korea, \{mjjeon, jyjung, jjl\}@maum.ai \\
         $^3$ Soongsil University, Seoul, Republic of Korea \\
         $^4$ Independent Researcher, tonyswoo@gmail.com
         }

\begin{document}

\ninept
\maketitle

\begin{sloppy}

\begin{abstract}
In this work, we aim to analyze and optimize the EnCLAP framework, a state-of-the-art model in automated audio captioning. We investigate the impact of modifying the acoustic encoder components, explore pretraining with different dataset scales, and study the effectiveness of a reranking scheme. Through extensive experimentation and quantitative analysis of generated captions, we develop EnCLAP++, an enhanced version that significantly surpasses the original.
\end{abstract}


\section{Introduction}
\label{sec:intro}
 Automated audio captioning (AAC), a cross-modal translation involving transcribing audio signals into concise and meaningful natural language descriptions \cite{aac}, remains a particularly challenging task with a substantial performance gap between human and machine. One significant contributor to the performance gap can be attributed to the intrinsic complexity of the task, as distinguishing between various sound events, especially between similar and ambiguous ones, requires extensive real-world knowledge. Furthermore, the scarcity of high-quality data, with the most widely used datasets, AudioCaps \cite{audiocaps} and Clotho \cite{clotho} containing only 50K and 20K captions, respectively, poses an additional challenge. To address these challenges, prior studies have employed pretrained audio encoders trained on audio classification tasks \cite{mei, conette, beats-conformer}, leveraged the text generation capabilities of pretrained language models like GPT-2 \cite{gpt2, prefix_tuning, pengi} and BART \cite{bart, gontier}, and incorporated auxiliary loss terms, including keyword prediction loss \cite{koizumi_keyword} or sentence embedding loss \cite{sentence_embedding}, to improve the semantic quality of captions and provide additional training signal.

Building on the previous line of research, Kim \etal \cite{enclap} proposed the EnCLAP framework which integrates a set of pretrained models with an auxiliary training task. Specifically, EnCLAP utilizes two acoustic feature encoders, EnCodec \cite{encodec} and CLAP \cite{clap_laion}, to generate timestep-level and sequence-level representation of the input audio sequence, respectively. EnCLAP utilizes pretrained BART as the caption decoder to leverage these features and generate captions. Furthermore, Kim \etal also introduced masked codec modeling (MCM), an auxiliary task which involves masking a part of the input codec sequence and predicting it, to enhance the acoustic awareness of the caption decoder. The caption decoder was trained jointly using cross-entropy loss for caption generation and MCM loss. The combination of these approaches allowed EnCLAP to achieve state-of-the-art performance on the AudioCaps dataset.

Although EnCLAP exhibits impressive performance, the study by Kim \etal lacks sufficient experimental evaluation for determining the optimal models for the model components. Notably, the authors do not investigate alternative sequence-level acoustic features beyond CLAP. Furthermore, for timestep-level acoustic features, while they demonstrate that discrete codec input outperforms continuous input, their analysis is restricted to a single setup using EnCodec, without exploring other options or configurations. Additionally, Kim \etal acknowledge the issue of overfitting in larger model variants but do not investigate the use of large-scale weakly-labeled datasets \cite{wavcaps, beats-conformer}, which contain noisy and model-generated captions. Therefore, the EnCLAP framework has potential for further optimization. 

In this work, we extend and optimize the EnCLAP framework through a comprehensive examination of its components. We explore alternative acoustic feature encoder components and assess their efficacy. We also investigate the impact of large-scale training incorporating weakly-labeled datasets on the framework's performance. Furthermore, we adopt a sampling-and-reranking approach \cite{beats-conformer} as an alternative to beam search decoding and evaluate its effectiveness. Finally, we conduct a qualitative analysis of the generated captions to examine the effects of each component on the outputs. Based on our findings, we present EnCLAP++, an improved version of the EnCLAP model that achieved second place in the DCASE2024 Challenge Task6. Figure 1 provides an overview of EnCLAP++.

\section{Experimental Design}
\label{sec:method}
\begin{figure*}[t]
  \centering
  \centerline{\includegraphics[width=0.9\textwidth]{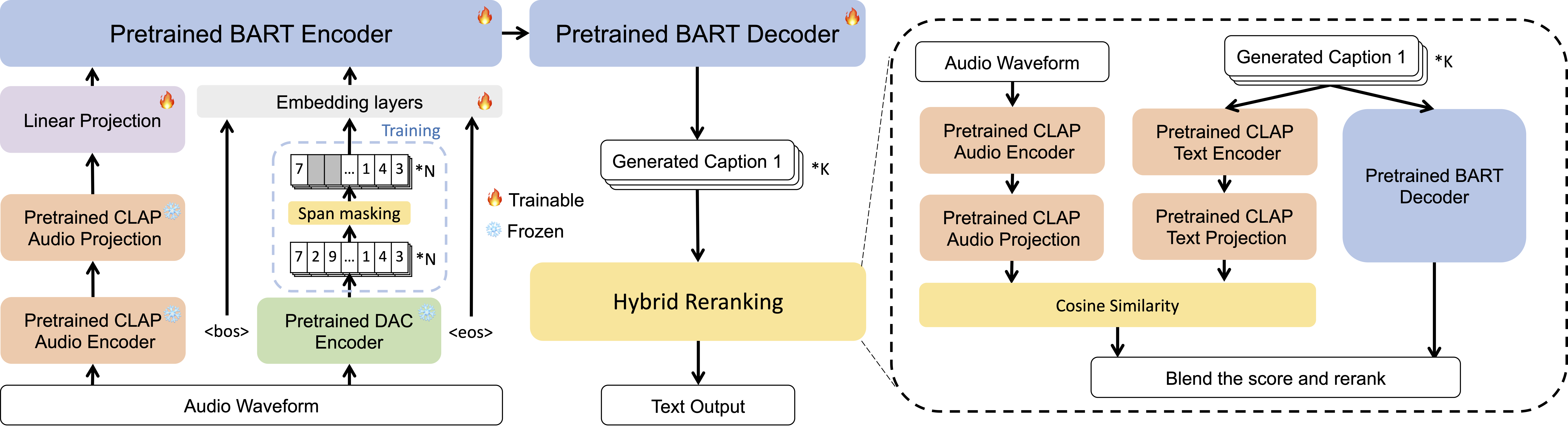}}
  \vspace{-3pt}
  \caption{Overall architecture of EnCLAP++}
  \label{fig:enclap++}
  \vspace{-10pt}
\end{figure*}

\subsection{Timestep-level Acoustic Embedding}
Neural audio codecs are autoencoder models designed to encode waveforms into sequences of discrete codes. 
Recent advancements \cite{soundstream, encodec, dac} typically employ residual vector quantization (RVQ) for compression, utilizing multiple codebooks to quantize the residuals of preceding codebooks. 
Ultimately, the input waveforms are transformed into a set of parallel discrete code sequences, each of which is associated with a unique codebook.
Neural audio codecs have demonstrated success as the acoustic representation format in generative audio models \cite{kreuk2022audiogen, valle, musicgen}. 

Kim \etal \cite{enclap} demonstrate that language models achieve superior performance when used with discrete input sequences compared to continuous input sequences. However, their study does not explore the impact of different configurations within the discrete input sequence setup. To address this limitation, we conduct experiments to examine the effects of different codec settings on the model performance. Specifically, we investigate the effect of codebook size, sample rate, and codec type on the final outcome.

The original EnCLAP employed a version of EnCodec \cite{encodec} that compresses a 24kHz audio signal into 16 discrete code sequences at a rate of 75Hz, with a codebook size of 1024. We experiment with two additional variants of EnCodec, which yield 8 and 32 code sequences, respectively, as well as a variant that processes 48kHz audio signal input. As for the alternative codec, we use a variant of Descript Audio Codec (DAC) \cite{dac} that closely resembles the original Encodec setup, which transforms 24kHz audio signal into 32 code sequences at a rate of 75Hz.
We opted for DAC as the alternative codec due to its superior performance in audio compression, as well as downstream tasks \cite{dac, codec_superb}.

\subsection{Sequence-level Acoustic Embedding} 
\label{subsec:clap}
While EnCLAP employs CLAP \cite{clap_laion} as its sequence-level acoustic feature encoder, preceding studies in audio captioning have predominantly utilized models pretrained on the AudioSet \cite{audioset} dataset for audio classification task \cite{mei, conette, beats-conformer}. In this work, we investigate alternative candidates for the sequence-level acoustic encoder component. Specifically, we examine the sequence-level representation capabilities of a model pretrained on AudioSet with audio tagging task and its variants, which have gone through additional audio-text retrieval training. We compare the audio captioning performance of these models with the original CLAP setup and assess the impact of additional retrieval training on downstream performance.

For the baseline sequence-level encoder, we use ConvNext-Tiny \cite{convnext} pretrained on AudioSet classification, referred to as CNext, and three of its variants that have undergone additional training on datasets of varying scales. Specifically, the three dataset configurations are: (1) Clotho \cite{clotho}, (2) AudioCaps \cite{audiocaps} and Clotho, and (3) WavCaps \cite{wavcaps}, AudioCaps, and Clotho. We use m-LTM framework \cite{m-ltm} and bge text encoder \cite{bge} for retrieval training. We assess the performance of these models against the original CLAP version. 

\begin{table*}{
\centering
\caption{Evaluation Results on Clotho. Ret refers to retrieval finetuning on the datasets listed in parentheses. CL, AC, and WC represent the Clotho, AudioCaps, and WavCaps datasets, respectively. Base and Large indicate the size of the pretrained BART model.}
\vspace{-0.5\baselineskip}
\label{table:ablation}
\begin{center}
\resizebox{0.8\textwidth}{!}{
\begin{tabular}{cccccccc}
\hline
Model & METEOR & CIDEr & SPICE & SPIDEr & SPIDEr-FL & Vocabulary & FENSE \\
\hline
\multicolumn{8}{c}{\textit{Timestep-level Representations}} \\
EnCLAP-base & 0.180 & 0.461 & 0.128 & 0.294 & 0.291 & 535 & 0.497 \\
w/ EnCodec, 8 codebooks &  0.178 &  0.444 & 0.127  & 0.286 & 0.283 & 626 & 0.497 \\ 
w/ EnCodec, 32 codebooks & 0.180  & 0.446 & 0.128  & 0.287  & 0.285 & \textbf{658} & 0.503 \\
w/ EnCodec, 48khz & 0.179 & 0.441 & 0.125 & 0.283 & 0.281 & 610 & \textbf{0.505} \\
w/ DAC & \textbf{0.183} & \textbf{0.463} & \textbf{0.131} & \textbf{0.297} & \textbf{0.294} & 589 & 0.504 \\
\cdashline{0-7}
\multicolumn{8}{c}{\textit{Sequence-level Representations}} \\
CLAP + DAC & \textbf{0.183} & \textbf{0.463} & \textbf{0.131} & \textbf{0.297} & \textbf{0.294} & 589 & 0.504 \\
CNext + DAC & 0.175 & 0.426 & 0.120 & 0.273 & 0.269 & 584 & 0.488 \\
CNext + Ret(CL) + DAC & 0.179 & 0.431 & 0.127 & 0.279 & 0.274 & \textbf{677} & 0.497 \\
CNext + Ret(CL+AC) + DAC & 0.181 & 0.454 & 0.130 & 0.292 & 0.287 & 596 & 0.500 \\
CNext + Ret(CL+AC+WC) + DAC & 0.179 & 0.452 & 0.127 & 0.290 & 0.286 & 676 & \textbf{0.508} \\
\cdashline{0-7}
\multicolumn{8}{c}{\textit{Large-Scale Pretraining}} \\
Base & 0.183 & 0.463 & 0.131 & 0.297 & 0.294 & 589 & 0.504 \\
Large & 0.184 & 0.393 & 0.132 & 0.262 & 0.260 & 571 & 0.480 \\
Base + WC Pretraining & 0.185 & \textbf{0.470} & \textbf{0.134} & \textbf{0.302} & \textbf{0.299} & \textbf{620} & \textbf{0.505} \\
Large + WC Pretraining & \textbf{0.187} & 0.464 & 0.130 & 0.297 & 0.293 & 576 & 0.500 \\
\cdashline{0-7}
\multicolumn{8}{c}{\textit{Generation and Reranking}} \\
Beam Search & 0.185 & 0.470 & 0.134  & 0.302 & 0.299  & 620  &  0.505 \\
Beam Search without Fluency error & 0.185 & 0.470 & 0.135 & 0.302 & 0.302 & 619 & 0.511 \\
Encoder Reranking & 0.176 & 0.396 & 0.126 & 0.261 & 0.261 & \textbf{915} & 0.520 \\
Decoder Reranking & 0.187 & 0.460 & 0.139 & 0.299 & 0.299 & 608 &  0.506 \\
Hybrid Reranking & \textbf{0.190}  & \textbf{0.479} & \textbf{0.142} & \textbf{0.310} & \textbf{0.310}  & 699 & \textbf{0.526} \\
\hline
\end{tabular}
}
\end{center}
\vspace{-1.2\baselineskip}
}\end{table*}

\subsection{Large-scale Pretraining}
The original EnCLAP described two versions of the model, denoted as "base" and "large", based on the size of the underlying BART \cite{bart} model used. The study highlights the issue of overfitting, especially in the large variant with smaller training datasets. To mitigate this issue, we draw on the recent trend in audio captioning, which involves leveraging weakly-labeled datasets for pretraining \cite{wavcaps, beats-conformer}. In particular, we evaluate a large-scale pretraining setup, where the model is pretrained on the WavCaps, and finetuned on Clotho, against the original EnCLAP dataset setup, where the model is pretrained on AudioCaps and finetuned on Clotho.
From WavCaps, we filter out audio clips that fall outside the 1-30 second duration range, as well as overlapping clips from AudioCaps and Clotho.
We evaluate both setups using both the base and large variants of our model. 

\subsection{Generation and Reranking}
Previous works, including EnCLAP, have utilized beam search decoding for caption generation. However, Wu \etal \cite{beats-conformer} demonstrates that the sampling-then-reranking approach yields more diverse and informative captions. Wu \etal proposes two scores for candidate reranking: the encoder reranking score and the decoder reranking score. The encoder reranking score is the cosine similarity score between the input audio representation and the generated caption representation computed using a retriever model. The decoder reranking score is the log-likelihood of the generated caption given the input audio. In this study, we explore the benefits of incorporating the reranking scheme into the EnCLAP framework. Specifically, we compare the original beam search scheme against three reranking setups: encoder reranking, decoder reranking, and hybrid reranking. 
We use CLAP as the retriever model for computing the encoder reranking score. 
We perform a fluency error-based filtering before the reranking procedure, following Wu \etal.

For sampling, we use nucleus sampling with a probability threshold of 0.95 and a temperature of 0.5 to generate 30 candidates. For hybrid reranking, we rank the candidates by the weighted sum of the encoder reranking score and the decoder reranking score using weights of 0.6 and 0.4, respectively.

\subsection{Quantitative Evaluation Metric} We adopt both widely used AAC metrics, METEOR, CIDEr, SPICE, and SPIDEr, and more recently proposed AAC metrics, SPIDEr-FL, FENSE \cite{fense}, and Vocab to evaluate various aspects of the generated captions. All metrics are calculated using the aac-metrics library. METEOR is a machine translation evaluation metric, based on unigram precision and recall. CIDEr and SPICE assess the syntactic and semantic quality of the generated captions, respectively, while SPIDEr is a linear combination of them. SPIDEr-FL is SPIDEr score penalized by the fluency error. FENSE is the combination of the SentenceBERT similarity score and the fluency error penalty. Vocab shows the diversity of the vocabularies in the generated captions.

\subsection{Qualitative Analysis}
Although quantitative metrics provide valuable insights into relative improvements in model performance, they are inherently limited, particularly in tasks such as audio captioning, where no single objective truth exists. Thus, in addition to reporting quantitative metrics, we perform a qualitative analysis of the generated captions. Specifically, we identify the examples with the largest improvement in the evaluation metric between the baseline and the best-performing variant and manually examine the enhancement in the caption quality.

\section{Results and Analysis}
\label{sec:experiment}
\begin{table*}
\centering
\caption{Example of the generated captions.}
\vspace{0.3\baselineskip}
\label{table:qualitative}
\resizebox{0.95\textwidth}{!}{
\begin{tabular}{ccc}
\hline
\multicolumn{3}{c}{\textit{Timestep-level Representations}} \\
 w/o DAC & w/ DAC & Ground Truth \\
\hline
A person walks on a hard & A person is walking on a hard surface & A person walking down a beach boardwalk with seagulls squawking \\
surface at a constant pace & while birds are chirping & overhead and people chatting in the background near the end \\
\hdashline
A woman is speaking over an & A man is speaking on a radio & A man is talking on a radio \\
intercom to a crowd of people & with people talking in the background & with singing in the background\\
\hdashline
A door creaks as it is opened & A person is walking on a wooden floor & Someone walking slowly \\
and closed several times &  while birds chirp in the background & as birds chirp in the background \\
\hline
\multicolumn{3}{c}{\textit{Sequence-level Representations}} \\
w/o CLAP & w/ CLAP & Ground Truth \\
\hline
Water is running from & A person is walking through & Someone is walking outside \\
a faucet into a sink & a pile of leaves & on a path covered with dried leaves \\
\hdashline
The wind is blowing and & A group of children are & Many children are talking and \\
a car is driving by & yelling and screaming & screaming, all at the same time \\
\hdashline
A heavy rain coming down & A saw is being used & A saw being used to saw wood that \\
outside during a storm & to cut a piece of wood & makes squeaking noises at the end\\
\hline
\multicolumn{3}{c}{\textit{Generation Scheme}} \\
Beam search & Reranking & Ground Truth \\
\hline
A gun is being fired & A hammer is repeatedly hit & Someone is repeatedly hitting \\
at a target & with a metal object & a hammer onto a wall or a nail \\
\hdashline
Birds are chirping and people & Children are playing, a car is driving, and & Children shout and play at the playground as \\
are talking in the background & birds are chirping & cars loudly drive by in the background \\
\hdashline
The engine of a car starts and & A motorcycle engine starts up and idles & A motorcycle engine starts \\
then the car drives away & for a while before idling down and idling again & and idles for a while \\
\hline
\end{tabular}
}
\end{table*}

\subsection{Timestep-level Acoustic Embedding}
Table \ref{table:ablation} shows that substituting the EnCodec encoder with an alternative variant does not enhance the model's performance and, in fact, leads to incremental degradation. This indicates that changing the timestep-level feature encoder across different EnCodec models has a negligible effect on the performance in the audio captioning task. Contrastively, replacing the EnCodec encoder with the DAC encoder leads to a modest improvement in the model performance. We believe that the DAC's superior ability to preserve the information in the original audio signal contributes to the enhancement. Therefore, we adopt DAC as the timestep-level acoustic feature encoder in subsequent experiments.

\subsection{Sequence-level Acoustic Embedding}
As illustrated in Table \ref{table:ablation}, the model using CNext as the sequence-level acoustic encoder falls behind the CLAP variant. However, the results indicate that additional retrieval training boosts the audio captioning performance and further, increasing the dataset size narrows the performance gap relative to the CLAP variant. Nevertheless, none of the CNext variants fully surpass the CLAP variant in terms of performance. We attribute the performance gap to the fact that CLAP was trained on a much larger scale than CNext, even with additional training, which is consistent with our findings within the CNext variants. Consequently, we will proceed with the original CLAP variant in subsequent experiments.

\subsection{Large-scale Pretraining}
\label{sec:ls}
The third section of Table \ref{table:ablation} demonstrates the effect of augmenting the pretraining dataset with a large-scale weakly-labeled dataset. Notably, our results for the original dataset setup replicate the phenomenon observed in the original EnCLAP work, where the large variant performs worse than the base variant. While variants with large-scale pretraining also exhibit this issue, the performance degradation is significantly less pronounced. Given that large-scale pretraining substantially improves the base variant, we infer that even the base variant can benefit from larger datasets. Our hypothesis is that larger datasets are necessary to fully utilize the capabilities of the large variant models.

\subsection{Generation and Reranking}
We investigated sampling and reranking techniques using the base variant pretrained on WavCaps from Sec \ref{sec:ls}. The results are presented in the last section of Table \ref{table:ablation}. Our findings indicate that encoder reranking enhances both the diversity of words and the semantic content of the generated captions. However, this improvement in semantic quality comes at the expense of syntactic quality. In contrast, decoder reranking alone yields results comparable to beam search, while when encoder and decoder reranking are combined, there is a significant improvement in semantic quality without any degradation in syntactic quality.

\subsection{Qualitative Analysis}
% We also conduct qualitative analysis of the generation captions.

\noindent\textbf{Timestep-level Acoustic Embedding. } The variant without DAC tends to focus on the most prominent event in a clip, but frequently overlooks background and supplementary acoustic events. This shortcoming can be attributed to the inherent constraint of relying on a single vector to represent the entire clip, which can lead to a loss of details. The inclusion of DAC, a timestep-level representation, enables the model to capture more fine-grained details of the scene.

\noindent\textbf{Sequence-level Acoustic Embedding. } While the model without CLAP generally succeeds in capturing the atmosphere of the acoustic scene, it tends to confound the overall semantic meaning of the scene. Thus, its captions describe an event similar to the actual event, but is actually different. We believe this comes from the lack of world knowledge to clear up the ambiguity. Thus, the variant with CLAP does not suffer from this issue. We attribute this to the model's lack of world knowledge, which fails to resolve ambiguities. Consequently, its generated captions describe an event that is similar to, yet distinct from, the actual event. In contrast, the variant with CLAP does not suffer from this issue.

\noindent\textbf{Generation and Reranking. } The captions produced by beam search variants are typically shorter and more concise, often omitting scene details. In contrast, the reranking variant generates more detailed captions that closely align with the label captions.

\begin{table}{
\centering
\vspace{-10pt}
\caption{Result on AudioCaps}
\vspace{-1.4\baselineskip}
\label{table:audiocaps}
\begin{center}
\resizebox{\linewidth}{!}
{
\begin{tabular}{cccccc}
\hline
Model & METEOR & CIDEr & SPICE & SPIDEr & FENSE \\
\hline
AL-MixGen~\cite{multitta} & 0.242 & 0.769 & 0.181 & 0.475 &  - \\
Wavcaps~\cite{wavcaps} & 0.250 & 0.787 & 0.182 & 0.485 & - \\
CoNeTTE~\cite{conette}  & 0.253 & 0.806 & 0.184 & 0.495 & 0.643 \\
EnCLAP-base~\cite{enclap} & 0.247 & 0.780 & 0.186 & 0.483 & 0.650 \\
EnCLAP-large~\cite{enclap} & 0.255 & 0.803 & 0.188 & 0.495 & 0.655 \\
\cdashline{0-5}
EnCLAP++-base& 0.257 & 0.815 & 0.188 & 0.501 & 0.661 \\
EnCLAP++-large & \textbf{0.269} & \textbf{0.823} & \textbf{0.197} &  \textbf{0.510} &\textbf{0.665} \\
\hline
\end{tabular}
}
\end{center}
\vspace{-1.3\baselineskip}
}\end{table}
\subsection{Results on AudioCaps}
Based on observations from Section \ref{sec:experiment}, we propose EnCLAP++, an improved version of EnCLAP that incorporates DAC, large-scale pretraining, and hybrid reranking. We evaluate EnCLAP++ on the AudioCaps dataset and present the results in Table \ref{table:audiocaps}. The assessment shows that both EnCLAP++-base and EnCLAP++-large outperform their respective EnCLAP counterparts, demonstrating the effectiveness of our mix of optimizations across different datasets.

\begin{table}{
\centering
\vspace{-10pt}
\caption{DCASE 2024 Challenge Result on Clotho Evaluation Split}
\vspace{-1.4\baselineskip}
\label{table:main}
\begin{center}
\resizebox{\linewidth}{!}
{
\begin{tabular}{cccccc}
\hline
Model & METEOR & CIDEr & SPICE & SPIDEr & FENSE \\
\hline
DCASE 2024 Baseline & 0.186 & 0.442 & 0.135 & 0.288 & 0.510 \\
Feng \etal \cite{kong_cuhk_t6_2024} & 0.192 & 0.495 & 0.141  & 0.318 & 0.525 \\
Kim \etal \cite{kyogu_snu_t6_2024} & 0.189 & 0.409 & 0.135 & 0.272 & 0.526 \\
Liu \etal \cite{li_alxc_t6_2024} & 0.195 & 0.493 & 0.145 & 0.319 & 0.533 \\
Chen \etal \cite{chen_sjtu_t6_2024} & 0.194 & \textbf{0.509} & 0.145 & \textbf{0.327} & 0.541 \\
Jung \etal \cite{jung_cmu_t6_2024} & 0.172 & 0.344 & 0.140 & 0.242 &  \textbf{0.554}  \\
\cdashline{0-5}
EnCLAP++ & \textbf{0.199} & 0.480 & \textbf{0.148} & 0.314 & 0.544 \\
\hline
\end{tabular}
}
\end{center}
\vspace{-1.2\baselineskip}
}\end{table}
\subsection{Results on DCASE Challenge 2024}
We submitted a variant of EnCLAP++ to the DCASE Challenge 2024. This variant employs a large version of BART and is pretrained on an extensive dataset that combines WavCaps, AudioCaps, and Clotho-Chatmix \cite{beats-conformer}. Due to the challenge regulations, we could not use CLAP because of potential overlap with the evaluation dataset. Therefore, we adopted CNext from Sec \ref{subsec:clap}, which was additionally trained with text-retrieval on WavCaps, AudioCaps, and Clotho, as the sequence-level representation.

The overall results are presented in Table \ref{table:main}. Our model achieved second place in the challenge, which was ranked based on the FENSE metric. Additionally, our model outperformed all other models on the METEOR and SPICE metrics.

\section{Conclusion}
\label{sec:conclusion}
This study presents an analysis of the EnCLAP framework and its components. Our investigation reveals that replacing the EnCodec encoder with the DAC encoder, augmenting the pretraining dataset with large-scale weakly-labeled data, and the incorporating of a reranking scheme enhances the model’s performance in audio captioning. Notably, our modified variant, EnCLAP++ shows significant improvement over the original model. Future directions for our research involve extending the EnCLAP framework to incorporate recent advances in large language models,  thereby enhancing its capabilities.
\bibliographystyle{IEEEtran}
\bibliography{main}

\end{sloppy}
\end{document}